\documentclass[preprint,aps,12pt,preprintnumbers,eqsecnum,nofootinbib]{revtex4}
\usepackage{graphicx}
\usepackage{subfigure}

\usepackage{color}
\usepackage{amssymb,amsmath}

\newcommand{\beq}{\begin{equation}}
\newcommand{\enq}{\end{equation}}

\DeclareMathOperator{\tr}{tr}
\begin{document}
%
%
\title{\vspace*{0.5in}

Dark chiral symmetry breaking \\ and the origin of the electroweak scale
\vskip 0.1in}
\author{Christopher D. Carone}\email[]{cdcaro@wm.edu}
\author{Raymundo Ramos}\email[]{raramos@email.wm.edu}

\affiliation{High Energy Theory Group, Department of Physics,
College of William and Mary, Williamsburg, VA 23187-8795}
\date{March 14, 2015}

\begin{abstract}
We study a classically scale-invariant model in which strong dynamics in a dark sector 
sets the scale of electroweak symmetry breaking.   Our model is distinct from others
of this type that have appeared in the recent literature.  We show that the Higgs sector of 
the model is phenomenologically viable and that the spectrum of dark sector states
includes a partially composite dark matter candidate.
\end{abstract}
\maketitle

\section{Introduction}
The Lagrangian of the standard model has precisely one dimensionful parameter, the squared mass of the Higgs doublet field.  This mass sets 
the scale of electroweak symmetry breaking, which is communicated to the standard model fermions via their Yukawa couplings.   The origin and
stability of the hierarchy between the electroweak scale and the Planck scale has motivated many of the leading proposals for physics beyond
the standard model. In this letter, we study the phenomenology of a specific model in which the Higgs mass squared arises as a result of strong 
dynamics in a dark sector.  Other models of this type have been discussed in the recent literature~\cite{strong1,strong2}; we explain how our model 
differs from those proposals below.

It is well known that the Yukawa coupling between a scalar $\phi$ and fermions can lead to a linear term in the scalar potential if the fermions 
condense.   Such a term alters the potential so that the scalar develops a vacuum expectation value (vev).  If the scalar squared mass term 
is absent, then the scale of the scalar vev is set entirely by that of the strong dynamics that produced the condensate.  If these fields carry 
electroweak quantum numbers, then electroweak symmetry will be spontaneously broken.  A simple model based on this idea was proposed by 
Carone and Georgi in Ref.~\cite{Carone:1993xc}.  In this letter, we consider a similar theory in which the scalar and fermions in question do not carry 
electroweak charges.   The vev of $\phi$ does not break electroweak symmetry, but provides an origin for the Higgs squared mass via the 
Higgs portal coupling $\lambda_p \phi^\dagger \phi H^\dagger H$.   As long as $\lambda_p$ has the appropriate sign, electroweak symmetry 
breaking is triggered at a scale set by the strong dynamics of the dark sector.

The choice of a classically scale-invariant scalar potential can be justified by various arguments. We place them in 
two categories:

{\bf 1.} {\em The model is tuned.}   Dimensionful parameters might not assume natural values as a consequence of the 
probability distribution over the string landscape, which is poorly understood.  If one takes this point of view, it is not unreasonable to consider 
extensions of the standard model that are designed to address its deficiencies (for example, extensions that provide for viable dark matter 
physics) that appear tuned but are parametrically simple and can be easily tested in experiment.  Our model is of this type and 
could easily be ruled out (or supported) by upcoming dark matter searches.

{\bf 2.} {\em  The model is not tuned.}  If there are no physical mass scales between the weak and Planck scales, then the only possible
source of a Higgs quadratic divergences is from the cut off of the theory.   Although field theoretic completions to low-energy effective 
theories lead generically to quadratic divergences proportional to the square of the cutoff~\cite{Tavares:2013dga}, this may not be the case for 
quantum gravitational physics at the Planck scale~\cite{Dubovsky:2013ira}.  As argued in Ref.~\cite{Altmannshofer:2014vra},  a spacetime 
description itself may break down at this scale and one's intuition based on quantum field theories may be flawed.  If one takes this point of view, 
it is not unreasonable to assume that a Higgs mass generated via dimensional transmutation in the infrared is only multiplicatively 
renormalized~\cite{bardeen} and to explore the phenomenological consequences.  A significant number of recent papers have adopted this 
perspective~\cite{strong1,strong2,weakone,weak}.

The model we propose has a dark sector SU(2)$_L\times$SU(2)$_R$ chiral symmetry that is spontaneously broken by a fermion condensate 
triggered by strong dynamics. An SU(2)$_D$ subgroup of the global symmetry is gauged, and the dark fermions have Yukawa couplings 
to a scalar that is a doublet under this gauge symmetry.  The dark sector would be an electroweak neutral clone of the model in 
Ref.~\cite{Carone:1993xc}, except that a U(1) gauge factor is replaced by a discrete subgroup to avoid a massless dark photon.  The presence of an 
SU(2)$_D$-doublet scalar immediately distinguishes the model from most related ones in the literature which employ a dark singlet to 
communicate dark sector strong dynamics through the Higgs portal~\cite{strong1}.  We note that the model of Ref.~\cite{strong2} has the same dark sector global chiral 
symmetry as ours, but does not gauge any subgroup.  This leads to a different particle spectrum and phenomenology.  We also utilize a non-linear 
chiral Lagrangian approach, familiar from the study of technicolor and QCD, which provides a convenient framework for the systematic 
description of dark sector phenomenology at low energies.

Our paper is organized as follows: In the next section we define the model.  In Sec.~3, we consider phenomenological
constraints.  In Sec.~4, we study the relic density and direct detection of the dark matter candidate in the model, which
is a partially composite dark sector state. In Sec.~5, we present our conclusions.

\section{The model} \label{sec:model}

The gauge group of the model is $G_{\rm SM} \times$ SU($N$)$\times$SU(2)$_D$.  The first factor refers to the
standard model gauge group, while the second is responsible for confinement in the dark sector. The
$G_{\rm SM}$ singlet fields (which we will call the dark sector, henceforth) are: a complex SU(2)$_D$-doublet scalar $\phi$, a 
left-handed SU(2)$_D$-doublet fermion $\Upsilon_L \equiv (p_L, m_L)^T$ and two right-handed singlet fermions $p_R$ and $m_R$. The fermions
transform in the fundamental representation of the  SU($N$) group. The field content is analogous to that of the technicolor model in 
Ref.~\cite{Carone:1993xc} with SU(2)$_W$ replaced by SU(2)$_D$ and U(1)$_Y$  replaced by a $Z_3$ factor.  As we will see below, the 
latter choice is the simplest way to preserve a convenient analogy between the two theories while also eliminating an unwanted massless gauge 
field. The dark sector has a global SU(2)$_L \times$SU(2)$_R$ chiral symmetry that is 
spontaneously broken when the dark fermions condense
\begin{equation}
\langle \overline{p} \, p + \overline{m} \, m \rangle \approx 4 \pi f^3 \,\,\, ,
\end{equation}
where $f$ is the dark pion decay constant.  We refer to the unbroken SU(2) subgroup of the global symmetry as dark isospin.
Spontaneous chiral symmetry breaking results in an isotriplet of dark pions
\begin{equation}
\Pi = \sum_{a=1}^3 \pi^a \frac{\sigma^a}{2} \,\,\, ,
\end{equation}
where $\sigma^a$ are the Pauli matrices.   As in the chiral
Lagrangian approach of Ref.~\cite{Carone:1993xc}, we adopt a nonlinear representation 
\begin{equation}
  \Sigma=\exp (2i\Pi/f) \,\,\, ,
\end{equation}
which transforms under the global chiral symmetry as $\Sigma\to L\Sigma R^\dagger$, where
$L$ and $R$ are the transformation matrices for SU$(2)_L$ and SU$(2)_R$, respectively.  It will be convenient to
define the following four-by-four matrix field
\begin{equation}
\Phi \equiv \left( \begin{array}{c|c} i \sigma^2 \phi^* &\,\,  \phi \end{array} \right)  \,\,\, ,
\end{equation}
and the nonlinear field redefinition 
\begin{equation}
\Phi= \frac{\sigma+f'}{\sqrt{2}} \Sigma'
\end{equation}
with $\Sigma'=\exp(2i\Pi'/f')$.
The kinetic terms for $\Phi$ and $\Sigma$ are
\begin{align}
  \mathcal{L}_{KE}&=\frac{1}{2}\tr\left(D_\mu \Phi^\dagger D^\mu \Phi\right)
    +\frac{f^2}{4}\tr\left(D_\mu \Sigma^\dagger D^\mu \Sigma\right)\nonumber\\
    &=\frac{1}{2}\partial _\mu \sigma \partial ^\mu \sigma
      +\frac{f^2}{4}\tr\left(D_\mu \Sigma^\dagger D^\mu \Sigma\right)
      +\frac{(\sigma+f')^2}{4}\tr\left(D_\mu \Sigma'^\dagger D^\mu \Sigma'\right)  \,\,\, .
\end{align}
Here $D_\mu=\partial_\mu-i g_D A^a_\mu\frac{\sigma^a}{2}$, where $A_\mu^a$ is the SU(2)$_D$ gauge field.
Study of the terms quadratic in the fields allows one to identify an unphysical linear combination of fields $\Pi_u$
that becomes the longitudinal component of $A^a_\mu$, and an orthogonal state $\pi_p$ that is physical:
\begin{equation}
  \pi_u=\frac{f\Pi+f'\Pi'}{\sqrt{f^2+f'^2}},
  \label{piona}
\end{equation}
\begin{equation}
  \pi_p=\frac{-f'\Pi+f\Pi'}{\sqrt{f^2+f'^2}}.
  \label{pionp}
\end{equation}
The $\pi_p$ multiplet will later be identified as the dark matter candidate in the theory.

Explicit breaking of the chiral symmetry originates from the Yukawa couplings.  Assuming that the fields transform under the 
$Z_3$ symmetry as
\begin{equation}
   \Upsilon_L\to\Upsilon_L \, ,  \:\:\:\:\:\: \phi\to\omega \, \phi \, ,  \:\:\:\:\:\:p_R\to\omega \, p_R \, , \:\:\:\:\:\: m_R\to\omega^2 \, m_R \, ,
\end{equation}
where $\omega^3=1$,  we find that the Yukawa couplings are given as in Ref.~\cite{Carone:1993xc} by
\begin{equation}
- \mathcal{L}_{y}= y_+ \overline{\Upsilon}_L \tilde{\phi} \, p_R+y_-\overline{\Upsilon}_L \phi  \,m_R  + \mbox{h.c.} \,\,\, .
\end{equation}
Defining $\Upsilon_R \equiv (p_R , m_R)$ and the matrix $Y \equiv \mbox{diag}(y_+ , y_-)$ this may be re-expressed as
\begin{equation}
- \mathcal{L}_{y} = \bar{\Upsilon}_L\Phi Y \Upsilon_R + \mbox{h.c.} \,\,\, ,
\end{equation}
which implies that we may treat $(\Phi Y)$ as a chiral-symmetry-breaking spurion with the transformation property
\begin{equation}
  (\Phi Y)\to L(\Phi Y) R^\dagger  \,\,\, .
\end{equation}
The lowest order term in the chiral Lagrangian that involves  $(\Phi Y)$ is
\begin{equation}
  \mathcal{L} = c_1 4\pi f^3 \tr (\Phi Y \Sigma^\dagger)+ \text{h.c.}
  \label{lspurion}
\end{equation}
where $c_1$ is expected to be of order unity by naive dimensional analysis~\cite{nda}.   This term determines the physical
dark pion mass
\begin{equation}
m_\pi^2 = 2 c_1 \sqrt{2} \frac{4 \pi f}{f'} (f^2+{f'}^2) \, y \,\,\, ,
\label{eq:pimass}
\end{equation}
where $y\equiv (y_+ + y_-)/2$, as well as a linear term in the scalar potential
\begin{equation}
V_y(\sigma)=-8 \sqrt{2} \pi c_1 f^3 y \, \sigma \,\,\, .
\label{linsig}
\end{equation}
This term sets the scale of the dark scalar vev, which determines the induced mass term for
the standard model Higgs doublet $H$ via a coupling in the potential  $V = V_0 + V_y$, where $V_0$
represents the scale-invariant terms:
\begin{equation}
  V_0(\phi,H)=\frac{\lambda}{2}(H^\dagger H)^2-\lambda_p(H^\dagger H)(\phi^\dagger\phi)+\frac{\lambda_\phi}{2} (\phi^\dag \phi)^2.
\label{eq:sipot}  
\end{equation}
In the ultraviolet (UV), before the dark fermions have condensed, vacuum stability of Eq.~(\ref{eq:sipot}) requires that
\begin{equation}
  \lambda>0\:\:\:\:\:\:\text{and}\:\:\:\:\:\:\lambda\lambda_\phi >\lambda_p^2.
\label{eq:stabcon}
\end{equation}
Noting that $\phi^\dagger \phi=\tr (\Phi^\dagger\Phi)/2=(\sigma+f')^2 /2$ and working in unitary gauge where $H = [0, (v+h)/\sqrt{2}]^T$,  the
potential may be re-expressed as
\begin{equation}
  V(h,\sigma)=\frac{\lambda}{8}(v+h)^4-\frac{\lambda_p}{4}(v+h)^2(\sigma+f')^2+\frac{\lambda_\phi}{8} (\sigma+f')^4-8\sqrt{2} \pi c_1 f^3 y \, \sigma  \,\,\,,
 \label{eq:lowpot} 
\end{equation}
after the dark fermions have condensed.  Minimization of  Eq.~(\ref{eq:lowpot}) leads to the following expressions for the vevs 
$v$ and $f'$: 
\begin{align}
  v^3 &= 2\left(\frac{\lambda_p}{\lambda}\right)^{3/2}
    \left( \lambda_\phi- \frac{\lambda_p^2}{\lambda}\right)^{-1}
    8 \sqrt{2}  \pi c_1f^3 y ,\label{eq:veq}\\
  f'^3 &= 2\left( \lambda_\phi- \frac{\lambda_p^2}{\lambda}\right)^{-1}
    8 \sqrt{2} \pi c_1  f^3 y.
    \label{eq:minc}
\end{align}
Of course, we fix $v = 246$~GeV to obtain the correct electroweak gauge boson masses.  The mass squared matrix in the
($h$, $\sigma$) basis is given by
\begin{equation}
  M^2=\left(\begin{array}{cc}
      \lambda & -\sqrt{\lambda\lambda_p}\\
      -\sqrt{\lambda\lambda_p}\:\:\: & \frac{1}{2}\left(\frac{3\lambda_\phi\lambda}{\lambda_p}-\lambda_p \right)
    \end{array}\right)v^2 \,\,\, ,
  \label{scmass}
\end{equation}
which is positive definite for positive couplings with  $\lambda\lambda_\phi >\lambda_p^2$.

One of the eingenvalues of this matrix corresponds to the squared mass of the
higgs scalar observed at the LHC, $m_{h_0}^2=(125.09\text{ GeV})^2$~\cite{Aad:2015zhl}. We call the remaining
mass eigenstate field $\eta$ below, and define the mixing angle $\theta$ by
\begin{equation} 
  \left(\begin{array}{cc}
    \cos\theta & -\sin\theta\\
    \sin\theta & \cos\theta
  \end{array}\right)
  \left(\begin{array}{c}
    h_0\\
    \eta
  \end{array}\right)=
  \left(\begin{array}{c}
    h\\
    \sigma
  \end{array}\right) \,\,\, .
\end{equation}
The value of the angle $\theta$ is given by $ \tan 2\theta=2 M^2_{12} / (M^2_{11}-M^2_{22})$
where $M^2_{jk}$ are elements of the matrix in Eq.~(\ref{scmass}). 

With the Higgs sector of the theory now defined, we proceed in the next section to study its phenomenology.   The parameters that define
the Higgs sector are $y_+$, $y_-$, $c_1$, $\lambda$, $\lambda_p$, $\lambda_\phi$, $f$, $f'$ and $v$.   We set the order-one coupling $c_1=1$ for
definiteness, and fix values for the Yukawa couplings assuming, for simplicity, that $y_+=y_-$.   The remaining six parameters are constrained by 
$v=246$~GeV, $m_{h_0}=125.09$~GeV, and the two minimization conditions given in Eqs.~(\ref{eq:minc})-(\ref{eq:veq}).  This leaves two degrees of freedom.  We choose 
the free parameters to be $f$ and $\lambda_p$ and map out the constraints on the model on the $f$-$\lambda_p$ plane.    This choice lends itself to
easy physical interpretation since $f$ parameterizes the scale of the dark sector strong dynamics, while $\lambda_p$ indicates how strongly the dark 
sector couples to the visible one.

\section{Phenomenological constraints}

We determine whether a given point on the $f$-$\lambda_p$ plane is allowed by imposing the following constraints:

1. {\em Absence of Landau poles below the Planck scale}.    The presence of such a Landau pole would suggest the onset of new
physics at an intermediate scale, contradicting our initial assumptions.  Since a coupling that blows up will become non-perturbative first, we 
eliminate the possibility of Landau poles by imposing perturbativity constraints on the running couplings.  For this purpose, we use the one-loop 
renormalization group equations (RGEs), which we provide in the appendix.  For the couplings $\lambda$, $\lambda_p$ and $\lambda_\phi$, one-loop 
corrections become equal in size to tree-level diagrams when, for example, $\lambda \approx 16 \pi^2$;  to avoid the complete breakdown of the perturbative
expansion, we set a generous upper limit on each of these couplings to be one-third of this value, $16 \pi^2 /3$, evaluated at all scales between
$m_Z$ and the reduced Planck mass $M_*$.   By similar reasoning, we place upper limits on the gauge and Yukawa couplings of $4\pi/\sqrt{3}$.  For our 
numerical results, we choose a perturbative value of the SU(2)$_D$ gauge coupling (35\% of $4\pi/\sqrt{3}$ in the example we present)  that is large enough to assure 
that the isotriplet gauge multiplet is heavier than the physical pions $\pi_p$; this will be required for our dark matter solutions, as discussed in the next section.  We take the  
SU($N$) gauge coupling to be at our perturbativity limit, $4 \pi /\sqrt{3}$, at $m_Z$ and choose $N=4$.   Since the  SU($N$) gauge coupling is asymptotically free in our 
theory, it remains perturbative for all scales higher than $m_Z$ (where we evaluate the RGEs), but it blows up quickly below $m_Z$, consistent with our assumption of 
strong dynamics in the infrared.

2. {\em Vacuum stability}.   The presence of the non-vanishing Higgs portal coupling requires that vacuum stability be studied in the
context of a two-Higgs doublet model.  In two-Higgs-doublet models, one can assure that the scalar potential remains bounded from below
by taking the stability conditions derived from the tree-level potential and testing whether they continue to hold for values of
the couplings evaluated at higher-renormalization scales, up to the Planck scale.  The justification for this approach can be found 
in Ref.~\cite{Sher:1988mj}.  We implement this by evaluating Eq.~(\ref{eq:stabcon})  using the one-loop renormalization group equations 
provided in the appendix. The scale at which a vacuum instability first arises depends on the free parameters of the model. A given point in 
the $f$-$\lambda_p$ plane satisfies the stability criterion if we find numerically that no violation of the stability conditions arises before the Planck scale.
For the dark-matter allowed points described later, the Higgs quartic coupling at the weak scale is larger than its standard model value, which
contributes to the model's vacuum stability.   

3. {\em Sufficiently standard-model-like Higgs boson}.   Standard model Higgs boson couplings are altered in this model by a factor of $\cos^2\theta$,
which can be no smaller that $0.7$ without spoiling global fits to Higgs data~\cite{cossq}.  The $\eta$ couplings to the visible sector are like those of the 
Higgs but suppressed by $\sin^2\theta$; non-observation of the $\eta$ in heavy Higgs search data from the LHC is assured for any $\eta$ mass within
the range experimentally studied, $145-710$~GeV, provided that $\sin^2\theta \lesssim 0.1$~\cite{hhs}. For simplicity, we require that each point in the 
$f$-$\lambda_p$ plane satisfy $\sin^2\theta < 0.1$.  Our final set of allowed points in parameter space discussed in Sec.~\ref{sec:dm} will 
correspond to $\eta$ masses in the range $178-203$~GeV, falling within the LHC range.  Note that we do not consider potentially tighter mixing angle 
bounds on very light $\eta$ from LEP2 since we will see later that this region of parameter space is excluded by our fourth constraint.

4. {\em Approximate chiral symmetry}.   Our effective chiral Lagrangian is valid provided that sources of explicit chiral symmetry breaking are small
compared to the chiral-symmetry-breaking scale $\Lambda_\chi \equiv 4 \pi f$.   We reject points in which the dark fermion masses $m_\pm$
exceed one-third $\Lambda_\chi$, or equivalently
\begin{equation}
\frac{1}{\sqrt{2}} \, y_\pm f' < \frac{4}{3} \pi f   \,\,\, .
\end{equation}
This assures that our initial assumption of an approximate SU(2)$_L \times$SU(2)$_R$ global symmetry remains valid.

\begin{figure}[t]
  \begin{center}
    \includegraphics[width=0.5\textwidth]{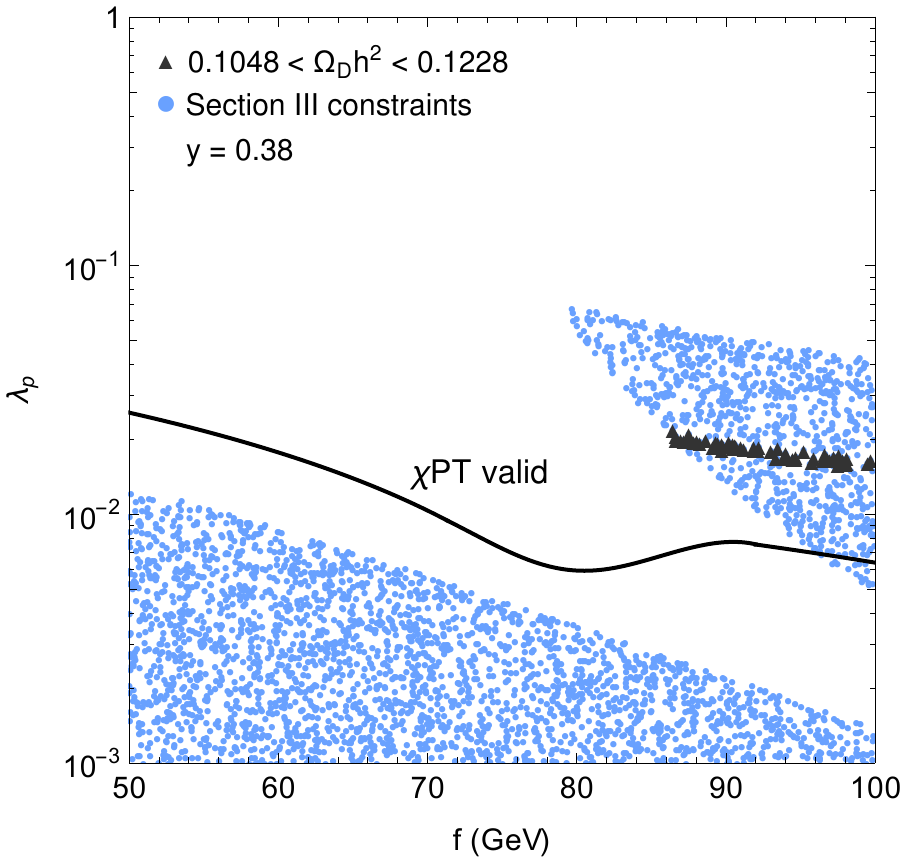}
    \caption{Regions of the parameter space consistent with perturbativity and stability constraints, as well as $\sin^2\theta<0.1$.  Points above the solid black line are
    consistent with approximate dark sector chiral symmetry.  Two branches of points correspond to $m_\eta > m_{h_0}$ (upper branch) and
    $m_\eta < m_{h_0}$ (lower branch).  The triangular points in the upper branch are consistent with current dark matter constraints.}   
        \label{fig:constraints}
  \end{center}
\end{figure}

We show results for a particular choice of $y$ in Fig.~\ref{fig:constraints}.   We have chosen to study values of $f$ near or below the scale where the
 SU($N$) gauge coupling becomes strong. The shaded regions satisfy the first three of the constraints discussed in this section.  The upper branch of points corresponds to 
an $\eta$ heavier that the SM Higgs boson, while the lower branch corresponds to the opposite.
The points which also satisfy our fourth constraint lie above the solid black line.    We find that viable dark matter solutions exist only for
$0.23 < y < 0.52$;  we have picked an intermediate value of $y$ as a representative choice. The dark matter 
results included in this figure will be discussed in the following section.

\section{Dark Matter} \label{sec:dm}

The dark sector of the model includes stable dark pions and baryons, provided that the pions are lighter than the baryons and the SU(2)$_D$ gauge multiplet.
In the case we consider, where  $y_+=y_-$, the stabilizing symmetry is the residual dark SU(2) isospin, which is non-anomalous and unbroken by higher-dimension 
operators (which are absent by the assumed scale invariance of the theory).   If $y_+$ and $y_-$ are unequal, then only the lightest of the dark pion triplet would be  stable; for 
simplicity, we consider the degenerate case here.  The dark baryons are separately stable due to a conserved dark baryon number.  However, estimating the dark baryon-anti-baryon 
annihilation cross section by scaling the analogous quantity measured experimentally in QCD,  we find that that dark baryon contribution to the relic density is orders of magnitude smaller 
than that of the $\pi_p$ for the parameter choices of relevance to our analysis\footnote{We will see in our figures that the relevant $\pi_p$ masses are comparable to the scale 
$\Lambda_\chi = 4 \pi f$, which we expect to be of order the dark baryon masses; however, in the effective chiral Lagrangian, the baryon mass terms involve additional unknown parameters 
that we may choose to assure that the dark baryons are heavier than the $\pi_p$.  We check directly that the SU(2)$_D$ gauge multiplet is also heavier.}.

The Higgs sector mixing angle $\theta$ is generally small, and we can estimate the annihilation cross section by the contributions that are 
lowest order in $\sin\theta$:  this selects $\pi_p^a \pi_p^a\to\eta\eta$, where 
$\pi_p = \sum_{a=1}^3\pi_p^a \sigma^a/2$, with $\pi_p$ defined in Eq.~(\ref{pionp}).  The $\pi_p \pi_p \eta$ and $\pi_p \pi_p h_0$ vertices
originate from Eq.~(\ref{lspurion}):
\begin{equation}
{\cal L} \supset -\frac{m_\pi^2}{2 \, f'} (\eta \cos\theta\ + h_0 \sin\theta) \, \pi_p^a \pi_p^a  \,\,\, .
\label{eq:pphe}
\end{equation}
The first term contributes to the annihilation process of interest via $t$- and $u$- channel pion exchange diagrams.  Working in the nonrelativistic limit, we find the thermally 
averaged annihilation cross section times velocity
\beq
  \langle\sigma_\text{ann} v \rangle = \frac{1}{16\pi}\frac{m_\pi^6}{f'^4}\left(1-\frac{m_\eta^2}{m_\pi^2}\right)^{1/2}\left[\frac{\cos^2\theta}{m_\eta^2-2m_\pi^2}\right]^2 \,\,\, ,
\enq
with $m^2_\pi$ given by Eq.~(\ref{eq:pimass}).   Using this, we calculate the freeze-out temperature $T_F$ and the dark matter relic density by standard methods~\cite{Kolb:1990vq}.  Defining
$x=m_\pi/T$ and taking into account the dark sector spectrum in evaluating $g_*(x)$, the number of relativistic degrees of freedom at the temperature $T$, we find
freeze-out temperatures near  $x_F\approx26$.  The relic density is given by
\beq
  \Omega_D h^2 \approx 3\cdot\frac{(1.07\times 10^9 \text{ GeV}^{-1})x_F}{\sqrt{g_*(x_F)}M_\text{Pl}\langle\sigma_\text{ann} v \rangle_F}
\label{relden}
\enq
which we require to reproduce the WMAP result $0.1138\pm0.0045$~\cite{WMAP} within two standard deviations.   In Fig.~\ref{fig:constraints}, the region consistent with $\pi_p$
dark matter is the band of triangular points in the upper branch of otherwise allowed points.  For our choice of $g_D \approx 2.54$, the SU(2)$_D$ gauge bosons are heavier than the $\pi_p$ for each 
triangular point shown.  We do not display results for other choices of $y$ in the range $0.23 < y < 0.52$ which are similar qualitatively to the plot in Fig.~\ref{fig:constraints}.   The main effect of 
increasing $y$ over this range is to enlarge the upper branch of points while moving the solid black exclusion line upwards until it is roughly contiguous with the band preferred by dark matter 
considerations when $y=0.52$. 

\begin{figure}[t]
  \begin{center}
    \includegraphics[width=0.5\textwidth]{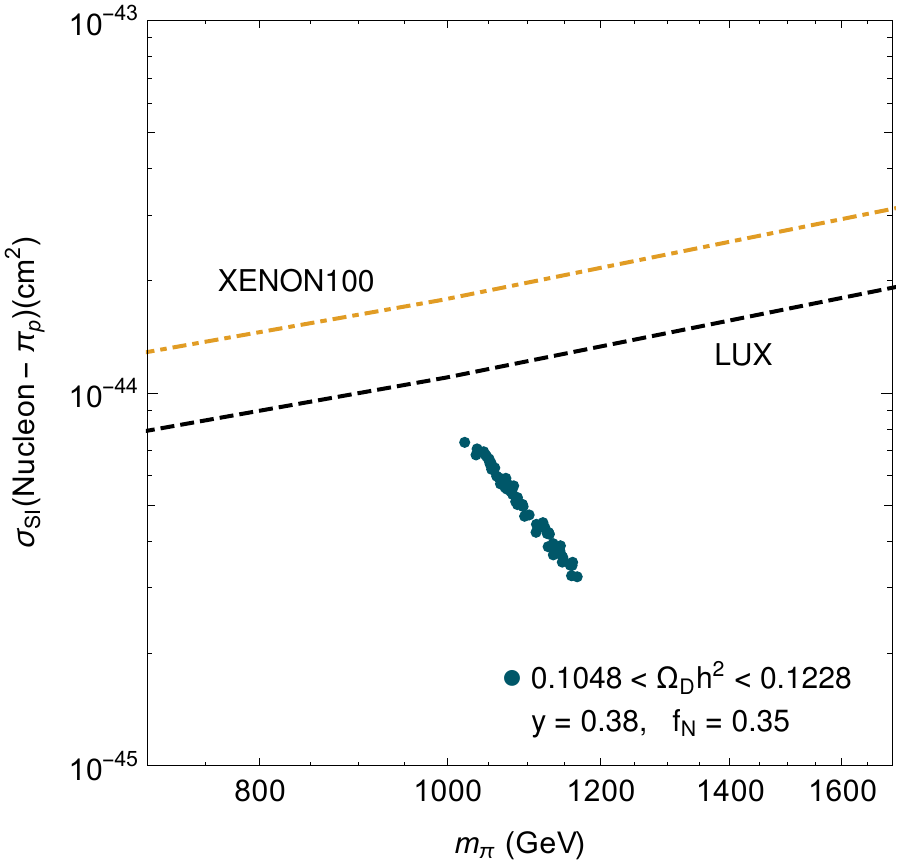}
    \caption{Dark pion-nucleon elastic scattering cross section for the points within the dark-matter-preferred band of Fig.~~\ref{fig:constraints}.   The current
    bounds from LUX~\cite{LUX} and XENON100~\cite{XENON100}  are also shown.}
    \label{fig:dirdec}
  \end{center}
\end{figure}

Finally, we compare the direct detection predictions of the model with current experimental bounds.  The $\pi_p$-nucleon spin-independent elastic scattering cross section is
determined by $t$-channel $h_0$ and $\eta$ exchange diagrams following from the vertices in Eq.~(\ref{eq:pphe}).   We find
\beq
  \sigma_{SI}(\pi_p N\to\pi_p N)=\frac{f_N^2}{16\pi}\frac{m_\pi^2m^2_N}{v^2f'^2}\sin^22\theta\frac{(m_\eta^2-m_{h_0}^2)^2}{m_\eta^4m_{h_0}^4}\left(\frac{m_N m_\pi}{m_N+m_\pi}\right)^2  \,\,\, ,
\enq
where $f_N$ parameterizes the Higgs-nucleon coupling and $m_N$ is the nucleon mass. The value of $f_N=0.35$ is used~\cite{fnref}.  Results corresponding to the dark-matter-preferred band
in Fig.~\ref{fig:constraints} are shown in Fig.~\ref{fig:dirdec}, which includes the current LUX~\cite{LUX} and XENON100~\cite{XENON100} bounds for comparison.  All the points shown are currently allowed by direct search constraints, though they are in a region not far from the current bounds.  This suggests that future results from the LUX experiment may begin to substantially restrict the preferred dark matter parameter space of the model.  

\section{Conclusions}

We have studied a classically scale-invariant model that provides an origin for the electroweak scale via dark sector strong dynamics.   The dark sector has a structure similar to
the bosonic technicolor model proposed in Ref.~\cite{Carone:1993xc}:  a fermion condensate is responsible for the instability that leads to a scalar doublet acquiring a vev.   In the model
of Ref.~\cite{Carone:1993xc}, the fermion condensate and the scalar vev each contribute to the breaking of electroweak symmetries. Here, the analogous fields are electroweak singlets;
the scalar vev breaks a dark SU(2) gauge group and induces a mass term for the standard model Higgs doublet field via couplings in the Higgs potential.  We found regions in the parameter 
space of the model where all the couplings can be run up to the Planck scale while remaining perturbative, where the scalar potential satisfies vacuum stability constraints, and 
where the Higgs boson is sufficiently standard-model-like to be consistent with existing collider data.  In addition, we showed that the partially composite dark isotriplet bosons
in the model can provide a viable dark matter candidate, providing the desired relic density while evading current direct detection bounds.  In addition, the model predicts that the dark matter-nucleon 
elastic scattering cross section lies just beyond the current LUX bounds.  Hence, the model may be ruled out, or given experimental support, as the LUX data set is enlarged.

\begin{acknowledgments}  
This work was supported by the NSF under Grant PHY-1068008.  
\end{acknowledgments}

\appendix*

\section{RGEs}
The RGEs used in our analysis are as follows:
\begin{align}
16 \pi ^2\frac{d\lambda_{\phi}}{dt} & =4 N \lambda_{\phi}
   \left(y_-^2+y_+^2\right)-4 N
   \left(y_-^4+y_+^4\right)-9
   g_D^2 \lambda_\phi+\frac{9}{4} g_D^4+4
   \lambda_p^2+12 \lambda_{\phi}^2  \,\, , \\
16 \pi ^2\frac{d\lambda}{dt} & =\frac{9}{4}\left(\frac{2}{5} g_1^2 g_2^2
   +\frac{3}{25}g_1^4+g_2^4\right)-\lambda \left(\frac{9}{5}g_1^2
	 +9 g_2^2\right)+12 \lambda
   y_t^2+12 \lambda^2+4 \lambda_p^2-12 y_t^4 \\
16 \pi ^2\frac{d\lambda_p}{dt} & = \left[2 N
   \left(y_-^2+y_+^2\right)+\frac{9}{2}
   \left(-\frac{1}{5}
   g_1^2-g_2^2-g_D^2\right)+6
   \lambda-4 \lambda_p+6 \lambda_{\phi}+6
   y_t^2\right]\lambda_p \,\, , \\
16 \pi ^2\frac{dy_t}{dt} & =\left[-\frac{17}{20}
   g_1^2-\frac{9}{4} g_2^2-8
   g_3^2+\frac{9}{2} y_t^2\right]y_t \,\, , \\
16 \pi ^2\frac{dy_-}{dt} & = \left[\left(N+\frac{3}{2}\right)
   y_-^2+\left(N-\frac{3}{2}\right)
   y_+^2-\frac{9}{4} g_D^2-\frac{3(N^2-1)}{N}g_{N}^2\right]y_- \,\, ,\\
16 \pi ^2\frac{dy_+}{dt} & =
   \left[\left(N-\frac{3}{2}\right)
   y_-^2+\left(N+\frac{3}{2}\right)
   y_+^2-\frac{9}{4} g_D^2-\frac{3(N^2-1)}{N} g_{N}^2\right] y_+ \,\, , \\
16 \pi ^2\frac{dg_D}{dt} & =\left[\frac{N}{3}-\frac{43}{6}
   \right] g_D^3 \,\, , \\
16 \pi ^2\frac{g_{N}}{dt} & =\left[\frac{4}{3}
	-\frac{11}{3}N\right]g_{N}^3 \,\, , \\
16 \pi ^2\frac{dg_i}{dt} & =b_ig_i^3 \,\,\, .
\end{align}
Here $t=\ln(\mu/m_Z)$, where $\mu$ is the renormalization scale, $b_i=\left( \frac{41}{10},-\frac{19}{6},-7 \right)$, the SU(5) normalization for the 
hypercharge was used and $g_N$ is the SU($N$) gauge coupling.

                                                                                                                                                          


\begin{thebibliography}{99}

\bibitem{strong1}
M.~Heikinheimo and C.~Spethmann,
  JHEP {\bf 1412}, 084 (2014)
  [arXiv:1410.4842 [hep-ph]];
J.~Kubo, K.~S.~Lim and M.~Lindner,
  JHEP {\bf 1409}, 016 (2014)
  [arXiv:1405.1052 [hep-ph]];
  O.~Antipin, M.~Redi and A.~Strumia,
  JHEP {\bf 1501}, 157 (2015)
  [arXiv:1410.1817 [hep-ph]];
  J.~Kubo, K.~S.~Lim and M.~Lindner,
  Phys.\ Rev.\ Lett.\  {\bf 113}, 091604 (2014)
  [arXiv:1403.4262 [hep-ph]];
  M.~Holthausen, J.~Kubo, K.~S.~Lim and M.~Lindner,
  JHEP {\bf 1312}, 076 (2013)
  [arXiv:1310.4423 [hep-ph]];
  M.~Heikinheimo, A.~Racioppi, M.~Raidal, C.~Spethmann and K.~Tuominen,
  Mod.\ Phys.\ Lett.\ A {\bf 29}, 1450077 (2014)
  [arXiv:1304.7006 [hep-ph]];
  T.~Hur and P.~Ko,
  Phys.\ Rev.\ Lett.\  {\bf 106}, 141802 (2011)
  [arXiv:1103.2571 [hep-ph]].
 
\bibitem{strong2}
T.~Hur, D.~W.~Jung, P.~Ko and J.~Y.~Lee,
  Phys.\ Lett.\ B {\bf 696}, 262 (2011)
  [arXiv:0709.1218 [hep-ph]].

\bibitem{Carone:1993xc} 
  C.~D.~Carone and H.~Georgi,
  Phys.\ Rev.\ D {\bf 49}, 1427 (1994)
  [hep-ph/9308205].
  
 \bibitem{Tavares:2013dga} 
  G.~Marques Tavares, M.~Schmaltz and W.~Skiba,
  Phys.\ Rev.\ D {\bf 89}, no. 1, 015009 (2014)
  [arXiv:1308.0025 [hep-ph]].

\bibitem{Dubovsky:2013ira} 
  S.~Dubovsky, V.~Gorbenko and M.~Mirbabayi,
  JHEP {\bf 1309}, 045 (2013)
  [arXiv:1305.6939 [hep-th]].

\bibitem{Altmannshofer:2014vra} 
  W.~Altmannshofer, W.~A.~Bardeen, M.~Bauer, M.~Carena and J.~D.~Lykken,
  JHEP {\bf 1501}, 032 (2015)
  [arXiv:1408.3429 [hep-ph]].
 
 \bibitem{bardeen}
W.~A.~Bardeen,
  FERMILAB-CONF-95-391-T, C95-08-27.3.
 
 \bibitem{weakone}
 R.~Hempfling,
  Phys.\ Lett.\ B {\bf 379}, 153 (1996)
  [hep-ph/9604278];
W.~-F.~Chang, J.~N.~Ng and J.~M.~S.~Wu,
  Phys.\ Rev.\ D {\bf 75}, 115016 (2007)
  [hep-ph/0701254 [HEP-PH]];
  R.~Foot, A.~Kobakhidze, K.~L.~McDonald and R.~R.~Volkas,
  Phys.\ Rev.\ D {\bf 76}, 075014 (2007)
  [arXiv:0706.1829 [hep-ph]];
 Phys.\ Rev.\ D {\bf 77}, 035006 (2008)
 [arXiv:0709.2750 [hep-ph]];
  T.~Hambye and M.~H.~G.~Tytgat,
  Phys.\ Lett.\ B {\bf 659}, 651 (2008)
  [arXiv:0707.0633 [hep-ph]];
  S.~Iso, N.~Okada and Y.~Orikasa,
  Phys.\ Lett.\ B {\bf 676}, 81 (2009)
  [arXiv:0902.4050 [hep-ph]];
   M.~Holthausen, M.~Lindner and M.~A.~Schmidt,
  Phys.\ Rev.\ D {\bf 82}, 055002 (2010)
  [arXiv:0911.0710 [hep-ph]];
  R.~Foot, A.~Kobakhidze and R.~R.~Volkas,
  Phys.\ Rev.\ D {\bf 82}, 035005 (2010)
  [arXiv:1006.0131 [hep-ph]];
  L.~Alexander-Nunneley and A.~Pilaftsis,
  JHEP {\bf 1009}, 021 (2010)
  [arXiv:1006.5916 [hep-ph]];
 K.~A.~Meissner and H.~Nicolai,
Phys.\ Lett.\ B {\bf 648}, 312 (2007)
[hep-th/0612165];
Phys.\ Lett.\ B {\bf 660}, 260 (2008)
[arXiv:0710.2840 [hep-th]];
S.~Iso and Y.~Orikasa,
  PTEP {\bf 2013}, 023B08 (2013)
  [arXiv:1210.2848 [hep-ph]];
 C.~Englert, J.~Jaeckel, V.~V.~Khoze and M.~Spannowsky,
  JHEP {\bf 1304}, 060 (2013)
  [arXiv:1301.4224 [hep-ph]].
  
 \bibitem{weak}
 J.~Guo, Z.~Kang, P.~Ko and Y.~Orikasa,
  arXiv:1502.00508 [hep-ph].
  H.~Okada and Y.~Orikasa,
  arXiv:1412.3616 [hep-ph].
 S.~Benic and B.~Radovcic,
  JHEP {\bf 1501}, 143 (2015)
  [arXiv:1409.5776 [hep-ph]].
  G.~M.~Pelaggi,
  Nucl.\ Phys.\ B {\bf 893}, 443 (2015)
  [arXiv:1406.4104 [hep-ph]].
 M.~Lindner, S.~Schmidt and J.~Smirnov,
  JHEP {\bf 1410}, 177 (2014)
  [arXiv:1405.6204 [hep-ph]];
 K.~Kannike, A.~Racioppi and M.~Raidal,
  JHEP {\bf 1406}, 154 (2014)
  [arXiv:1405.3987 [hep-ph]].
 A.~Farzinnia and J.~Ren,
  Phys.\ Rev.\ D {\bf 90}, no. 1, 015019 (2014)
  [arXiv:1405.0498 [hep-ph]];
 C.~Tamarit,
  Phys.\ Rev.\ D {\bf 90}, no. 5, 055024 (2014)
  [arXiv:1404.7673 [hep-ph]];
 K.~Allison, C.~T.~Hill and G.~G.~Ross,
  Phys.\ Lett.\ B {\bf 738}, 191 (2014)
  [arXiv:1404.6268 [hep-ph]];
 H.~Davoudiasl and I.~M.~Lewis,
  Phys.\ Rev.\ D {\bf 90}, no. 3, 033003 (2014)
  [arXiv:1404.6260 [hep-ph]];
 V.~V.~Khoze, C.~McCabe and G.~Ro,
  JHEP {\bf 1408}, 026 (2014)
  [arXiv:1403.4953 [hep-ph], arXiv:1403.4953];
 S.~Benic and B.~Radovcic,
  Phys.\ Lett.\ B {\bf 732}, 91 (2014)
  [arXiv:1401.8183 [hep-ph]];
 M.~Hashimoto, S.~Iso and Y.~Orikasa,
  Phys.\ Rev.\ D {\bf 89}, no. 5, 056010 (2014)
  [arXiv:1401.5944 [hep-ph]];
 J.~Guo and Z.~Kang,
  arXiv:1401.5609 [hep-ph];
 C.~T.~Hill,
  Phys.\ Rev.\ D {\bf 89}, no. 7, 073003 (2014)
  [arXiv:1401.4185 [hep-ph]];
 S.~Abel and A.~Mariotti,
  Phys.\ Rev.\ D {\bf 89}, no. 12, 125018 (2014)
  [arXiv:1312.5335 [hep-ph]];
M.~Hashimoto, S.~Iso and Y.~Orikasa,
  Phys.\ Rev.\ D {\bf 89}, no. 1, 016019 (2014)
  [arXiv:1310.4304 [hep-ph]];
T.~G.~Steele, Z.~W.~Wang, D.~Contreras and R.~B.~Mann,
  Phys.\ Rev.\ Lett.\  {\bf 112}, no. 17, 171602 (2014)
  [arXiv:1310.1960 [hep-ph]];
E.~Gabrielli, M.~Heikinheimo, K.~Kannike, A.~Racioppi, M.~Raidal and C.~Spethmann,
  Phys.\ Rev.\ D {\bf 89}, no. 1, 015017 (2014)
  [arXiv:1309.6632 [hep-ph]];
V.~V.~Khoze,
  JHEP {\bf 1311}, 215 (2013)
  [arXiv:1308.6338 [hep-ph]];
 A.~Farzinnia, H.~J.~He and J.~Ren,
  Phys.\ Lett.\ B {\bf 727}, 141 (2013)
  [arXiv:1308.0295 [hep-ph]];
  T.~Hambye and A.~Strumia,
  Phys.\ Rev.\ D {\bf 88}, 055022 (2013)
  [arXiv:1306.2329 [hep-ph]];
  C.~D.~Carone and R.~Ramos,
  Phys.\ Rev.\ D {\bf 88}, 055020 (2013)
  [arXiv:1307.8428 [hep-ph]];
  P.~Humbert, M.~Lindner and J.~Smirnov,
  arXiv:1503.03066 [hep-ph];
   K.~Endo and Y.~Sumino,
  arXiv:1503.02819 [hep-ph].
\bibitem{nda}
A.~Manohar and H.~Georgi,
  Nucl.\ Phys.\ B {\bf 234}, 189 (1984);
H.~Georgi and L.~Randall,
  Nucl.\ Phys.\ B {\bf 276}, 241 (1986).
    
\bibitem{Aad:2015zhl} 
  G.~Aad {\it et al.}  [ATLAS and CMS Collaborations],
  arXiv:1503.07589 [hep-ex].
  
  
\bibitem{Sher:1988mj} 
E.~Gildener and S.~Weinberg,
  Phys.\ Rev.\ D {\bf 13}, 3333 (1976);
  M.~Sher,
  Phys.\ Rept.\  {\bf 179}, 273 (1989);
  S.~Nie and M.~Sher,
  Phys.\ Lett.\ B {\bf 449}, 89 (1999)
  [hep-ph/9811234].
  
  
  \bibitem{cossq}
  J.~R.~Espinosa, C.~Grojean, M.~Muhlleitner and M.~Trott,
  JHEP {\bf 1212}, 045 (2012)
  [arXiv:1207.1717 [hep-ph]].
  
  \bibitem{hhs}
  S.~Chatrchyan {\it et al.}  [CMS Collaboration],
  Eur.\ Phys.\ J.\ C {\bf 73}, 2469 (2013)
  [arXiv:1304.0213 [hep-ex]].
  
  \bibitem{Kolb:1990vq} 
  E.~W.~Kolb and M.~S.~Turner,
  Front.\ Phys.\  {\bf 69}, 1 (1990).
  
  \bibitem{WMAP}
  C.~L.~Bennett {\it et al.}  [WMAP Collaboration],
  Astrophys.\ J.\ Suppl.\  {\bf 208}, 20 (2013)
  [arXiv:1212.5225 [astro-ph.CO]].

  \bibitem{fnref}
  J.~R.~Ellis, A.~Ferstl and K.~A.~Olive,
  Phys.\ Lett.\ B {\bf 481}, 304 (2000)
  [hep-ph/0001005].
  
\bibitem{LUX}
 D.~S.~Akerib {\it et al.}  [LUX Collaboration],
  Phys.\ Rev.\ Lett.\  {\bf 112}, 091303 (2014)
  [arXiv:1310.8214 [astro-ph.CO]].
  
\bibitem{XENON100}
 E.~Aprile {\it et al.}  [XENON100 Collaboration],
  Phys.\ Rev.\ Lett.\  {\bf 109}, 181301 (2012)
  [arXiv:1207.5988 [astro-ph.CO]].
  
\end{thebibliography}
\end{document}